\documentstyle[12pt,epsfig]{article}

\newcommand{\agt}{\,\rlap{\lower 3.5 pt \hbox{$\mathchar \sim$}} \raise 1pt
 \hbox {$>$}\,}
\newcommand{\alt}{\,\rlap{\lower 3.5 pt \hbox{$\mathchar \sim$}} \raise 1pt
 \hbox {$<$}\,}
\newcommand{\Arcosh}{\mathop{\mbox{Arcosh}}\nolimits}

\catcode`@=11
\newcount\@tempcntc
\def\@citex[#1]#2{\if@filesw\immediate\write\@auxout{\string\citation{#2}}\fi
  \@tempcnta\z@\@tempcntb\m@ne\def\@citea{}\@cite{\@for\@citeb:=#2\do
    {\@ifundefined
       {b@\@citeb}{\@citeo\@tempcntb\m@ne\@citea\def\@citea{,}{\bf
?}\@warning
       {Citation `\@citeb' on page \thepage \space undefined}}%
    {\setbox\z@\hbox{\global\@tempcntc0\csname b@\@citeb\endcsname\relax}%
     \ifnum\@tempcntc=\z@ \@citeo\@tempcntb\m@ne
       \@citea\def\@citea{,}\hbox{\csname b@\@citeb\endcsname}%
     \else
      \advance\@tempcntb\@ne
      \ifnum\@tempcntb=\@tempcntc
      \else\advance\@tempcntb\m@ne\@citeo
      \@tempcnta\@tempcntc\@tempcntb\@tempcntc\fi\fi}}\@citeo}{#1}}
\def\@citeo{\ifnum\@tempcnta>\@tempcntb\else\@citea\def\@citea{,}%
  \ifnum\@tempcnta=\@tempcntb\the\@tempcnta\else
   {\advance\@tempcnta\@ne\ifnum\@tempcnta=\@tempcntb \else
\def\@citea{--}\fi
    \advance\@tempcnta\m@ne\the\@tempcnta\@citea\the\@tempcntb}\fi\fi}
\catcode`@=12

\begin{document}
\title{
\vskip-3cm{\baselineskip14pt
\centerline{\normalsize DESY 01-039\hfill ISSN 0418-9833}
\centerline{\normalsize hep-ph/0104044\hfill}
\centerline{\normalsize April 2001\hfill}}
\vskip1.5cm
$J/\psi$ plus dijet associated production in two-photon collisions}
\author{{\sc M. Klasen, B.A. Kniehl, L. Mihaila, M. Steinhauser}\\
{\normalsize II. Institut f\"ur Theoretische Physik, Universit\"at
Hamburg,}\\
{\normalsize Luruper Chaussee 149, 22761 Hamburg, Germany}}

\date{}

\maketitle

\thispagestyle{empty}

\begin{abstract}
We study the production of a $J/\psi$ meson in association with one or two
jets in $\gamma\gamma$ collisions concentrating on the direct-photon
contribution, which is expected to be dominant for large $J/\psi$-meson
transverse momentum and/or large dijet invariant mass.
We work at leading order in the factorization formalism of nonrelativistic QCD
and include all relevant colour-octet processes.
We present distributions in $J/\psi$-meson transverse momentum and rapidity
appropriate for CERN LEP2, a future $e^+e^-$ linear collider, and a possible
$\gamma\gamma$ collider mode of the latter.
In the case of the $e^+e^-$ linear collider, we assume the beamstrahlung
spectrum appropriate for DESY TESLA.

\medskip

\end{abstract}

\newpage

\section{Introduction}

Since its discovery in 1974, the $J/\psi$ meson has provided a useful
laboratory for quantitative tests of quantum chromodynamics (QCD) and, in
particular, of the interplay of perturbative and nonperturbative phenomena.
The factorization formalism of nonrelativistic QCD (NRQCD) \cite{bbl} provides
a rigorous theoretical framework for the description of heavy-quarkonium
production and decay.
This formalism implies a separation of short-distance coefficients, which can 
be calculated perturbatively as expansions in the strong-coupling constant
$\alpha_s$, from long-distance matrix elements (MEs), which must be extracted
from experiment.
The relative importance of the latter can be estimated by means of velocity
scaling rules, i.e.\ the MEs are predicted to scale with a definite power of
the heavy-quark ($Q$) velocity $v$ in the limit $v\ll1$.
In this way, the theoretical predictions are organized as double expansions in
$\alpha_s$ and $v$.
A crucial feature of this formalism is that it takes into account the complete
structure of the $Q\overline{Q}$ Fock space, which is spanned by the states
${}^{2S+1}L_J^{(a)}$ with definite spin $S$, orbital angular momentum $L$,
total angular momentum $J$, and colour multiplicity $a=\underline{1}$,
$\underline{8}$.
In particular, this formalism predicts the existence of colour-octet processes
in nature.
This means that $Q\overline{Q}$ pairs are produced at short distances in
colour-octet states and subsequently evolve into physical (colour-singlet)
quarkonia by the nonperturbative emission of soft gluons.
In the limit $v\to 0$, the traditional colour-singlet model (CSM)
\cite{ber,bai} is recovered.
The greatest triumph of this formalism was that it was able to correctly 
describe \cite{yua,cho} the cross section of inclusive charmonium
hadroproduction measured in $p\overline{p}$ collisions at the Fermilab
Tevatron \cite{abe}, which had turned out to be more than one order of
magnitude in excess of the theoretical prediction based on the CSM.

In order to convincingly establish the phenomenological significance of the
colour-octet processes, it is indispensable to identify them in other kinds of
high-energy experiments as well.
Studies of charmonium production in $ep$ photoproduction, $ep$ and $\nu N$
deep-inelastic scattering, $e^+e^-$ annihilation, $\gamma\gamma$ collisions,
and $b$-hadron decays may be found in the literature; see Ref.~\cite{bra} and
references cited therein.
Furthermore, the polarization of charmonium, which also provides a sensitive
probe of colour-octet processes, was investigated \cite{ben,bkv,bkl,fle}.
None of these studies was able to prove or disprove the NRQCD factorization
hypothesis.
On the one hand, the theoretical predictions to be compared with existing
experimental data are, apart from very few exceptions \cite{kra,man,pet}, of
lowest order (LO) and thus suffer from considerable uncertainties, mostly from
the dependences on the renormalization and factorization scales and from the
lack of information on the nonperturbative MEs.
On the other hand, the experimental errors are still rather sizeable.
The latter will be dramatically reduced with the upgrades of DESY HERA and the
Fermilab Tevatron and with the advent of CERN LHC and hopefully a future
$e^+e^-$ linear collider (LC) such as DESY TESLA.
On the theoretical side, it is necessary to calculate the
next-to-leading-order (NLO) corrections to the hard-scattering cross sections
and to include the effective operators which are suppressed by higher powers
in $v$.

In this paper, we take a first step towards the complete NLO analysis of a
$2\to2$ process of heavy-quarkonium production in the NRQCD factorization 
formalism.
We consider the inclusive production of $J/\psi$ mesons in $\gamma\gamma$
collisions, $\gamma\gamma\to J/\psi+X$, where $X$ denotes the hadronic 
remnant.
The photons can originate from hard initial-state bremsstrahlung in $e^+e^-$
annihilation.
At high-energy $e^+e^-$ LCs, an additional source of hard photons is provided
by beamstrahlung, the synchrotron radiation emitted by one of the colliding
bunches in the field of the opposite bunch.
The highest possible photon energies with large enough luminosity may be
achieved by converting the $e^+e^-$ LC into a $\gamma\gamma$ collider via
back-scattering of high-energetic laser light off the electron and positron
beams.

The photons can interact either directly with the quarks participating in the
hard-scattering process (direct photoproduction) or via their quark and gluon
content (resolved photoproduction).
Thus, the process $\gamma\gamma\to J/\psi+X$ receives contributions from the
direct, single-resolved, and double-resolved channels.
All three contributions are formally of the same order in the perturbative
expansion.
This may be understood by observing that the parton density functions (PDFs)
of the photon have a leading behaviour proportional to
$\alpha\ln(M^2/\Lambda_{\rm QCD}^2)\propto\alpha/\alpha_s$, where $\alpha$ is
the fine-structure constant, $M$ is the factorization scale, and
$\Lambda_{\rm QCD}$ is the asymptotic scale parameter of QCD.
However, the direct channel is expected to be dominant at large $J/\psi$-meson
transverse momentum ($p_T$) and/or large invariant mass of the system $X$.
In the following, we focus our attention on the direct channel.

\begin{figure}[ht]
\begin{center}
\begin{tabular}{c}
\parbox{\textwidth}{\epsfig{figure=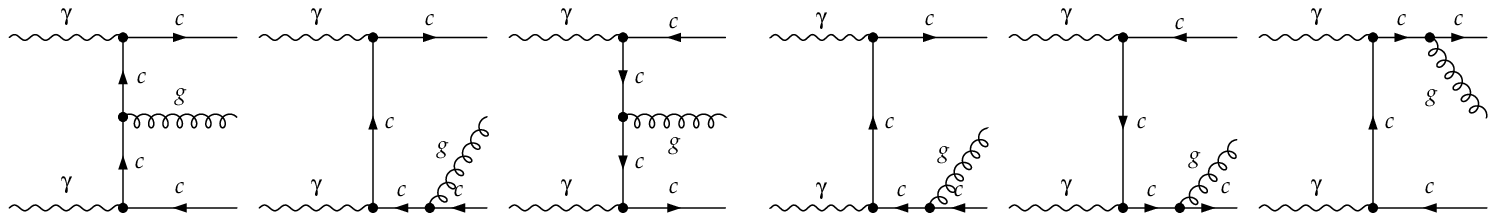,width=\textwidth}} \\
(a) \\
\parbox{\textwidth}{\epsfig{figure=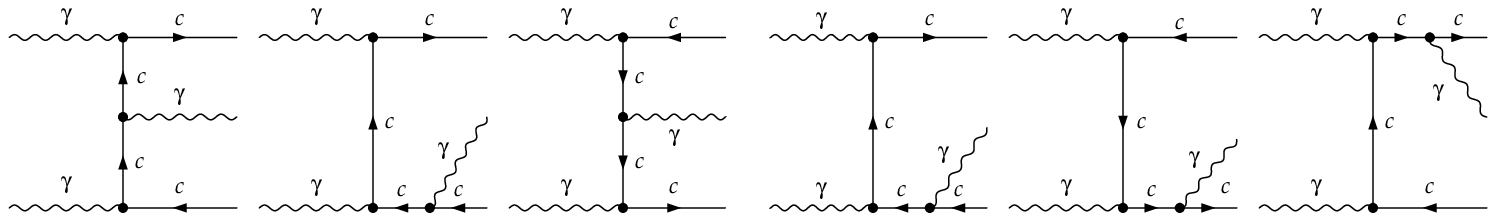,width=\textwidth}} \\
(b)
\end{tabular}
\caption{Feynman diagrams pertinent to the partonic subprocesses (a)
$\gamma\gamma\to c\bar cg$ and (b) $\gamma\gamma\to c\bar c\gamma$.}
\label{fig:ccX}
\end{center}
\end{figure}

\begin{figure}[ht]
\begin{center}
\begin{tabular}{c}
\parbox{\textwidth}{\epsfig{figure=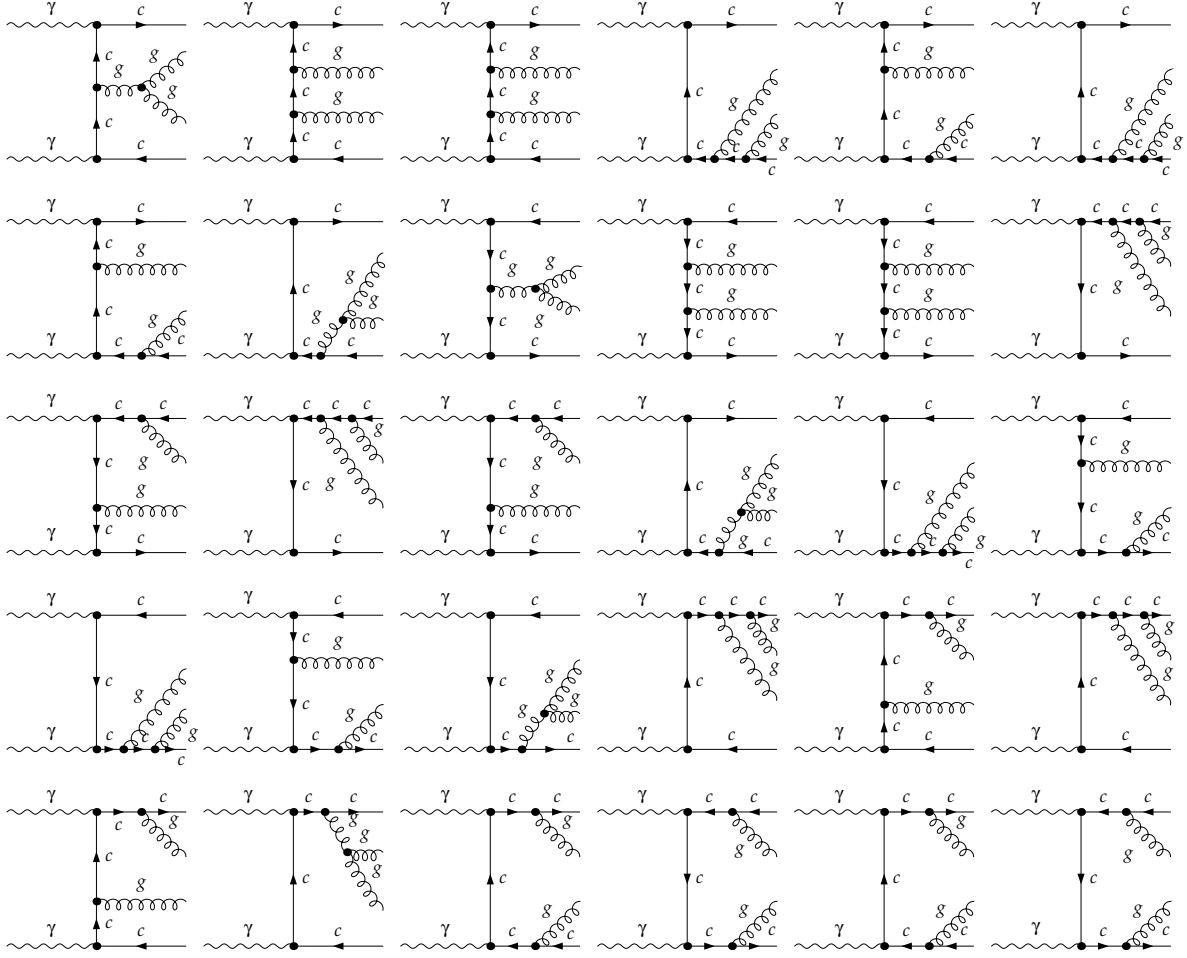,width=\textwidth}} \\
(a) \\
\parbox{\textwidth}{\epsfig{figure=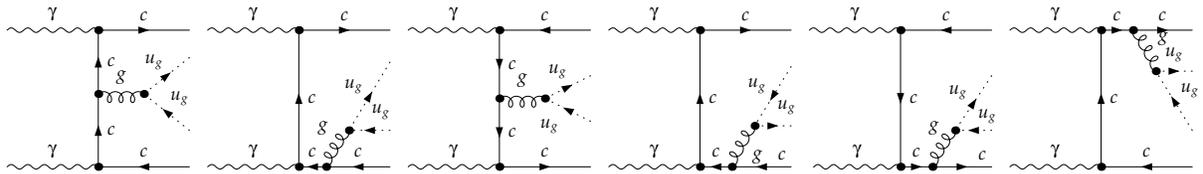,width=\textwidth}} \\
(b)
\end{tabular}
\caption{Feynman diagrams pertinent to the partonic subprocesses (a)
$\gamma\gamma\to c\bar cgg$ and (b) $\gamma\gamma\to c\bar cu_g\bar u_g$,
where $u_g$ and $\bar u_g$ are the Faddeev-Popov ghosts associated with the
gluon.}
\label{fig:ccgg}
\end{center}
\end{figure}

\begin{figure}[ht]
\begin{center}
\epsfig{figure=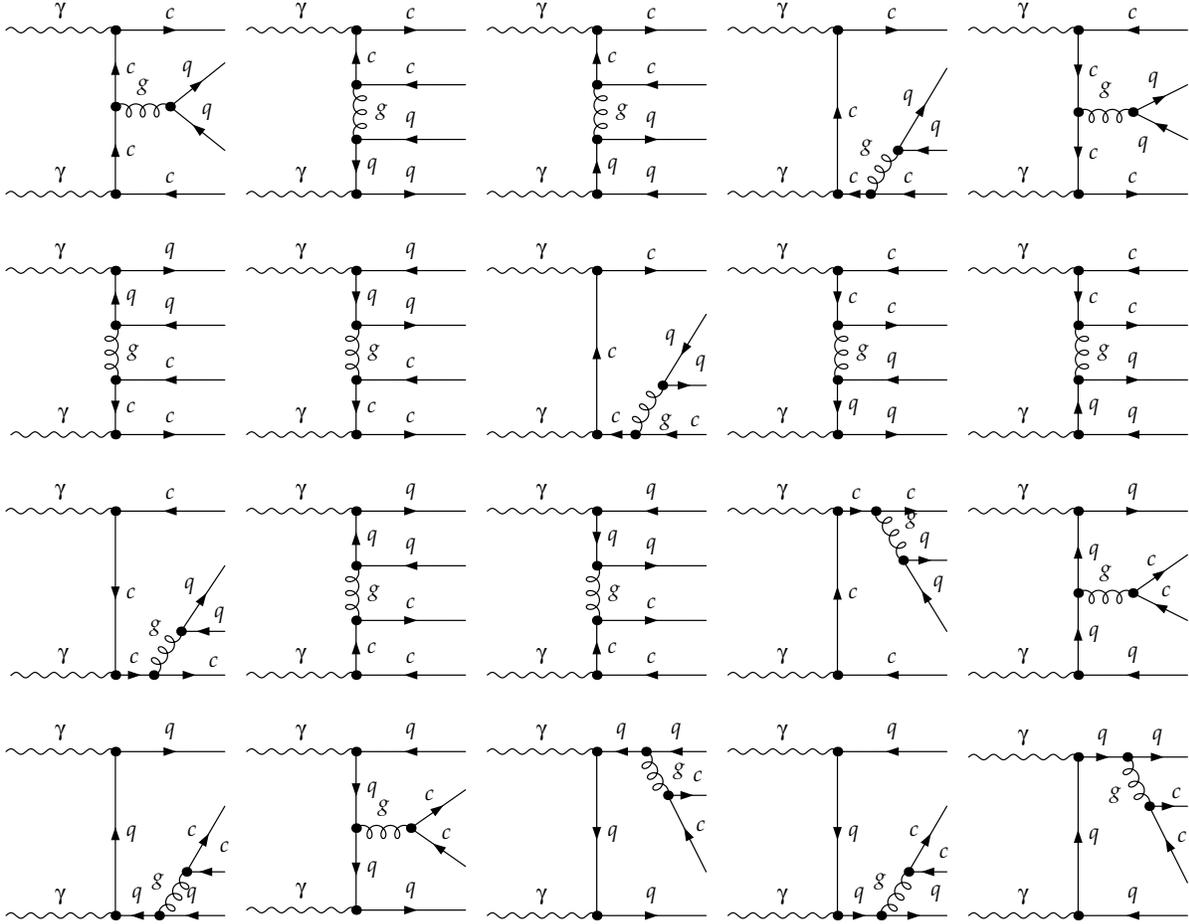,width=\textwidth}
\caption{Feynman diagrams pertinent to the partonic subprocess
$\gamma\gamma\to c\bar cq\bar q$, where $q=u,d,s$.}
\label{fig:ccqq}
\end{center}
\end{figure}

Let us consider the relevant partonic subprocesses that lead to a $J/\psi$
meson with finite $p_T$, i.e.\ we leave aside $2\to1$ processes.
The leading colour-singlet ME of the $J/\psi$ meson is
$\left\langle{\cal O}_1^{J/\psi}\left({}^3\!S_1\right)\right\rangle$, its
leading colour-octet ones are
$\left\langle{\cal O}_8^{J/\psi}\left({}^3\!S_1\right)\right\rangle$,
$\left\langle{\cal O}_8^{J/\psi}\left({}^1\!S_0\right)\right\rangle$, and
$\left\langle{\cal O}_8^{J/\psi}\left({}^3\!P_J\right)\right\rangle$, with
$J=0,1,2$.
If we restrict ourselves to purely hadronic final states, then the leading
$2\to2$ process is $\gamma\gamma\to c\bar c\left({}^3\!S_1^{(8)}\right)g$
\cite{ma,jap,god}.
The relevant Feynman diagrams are depicted in Fig.~\ref{fig:ccX}(a).
A colour-singlet process is only possible if the system produced together with
the $c\bar c$ pair forms a colour singlet, too.
Among the $2\to2$ processes, the leading such process is
$\gamma\gamma\to c\bar c\left({}^3\!S_1^{(1)}\right)\gamma$ \cite{ma,jap,god},
which proceeds through the Feynman diagrams shown in Fig.~\ref{fig:ccX}(b).
The leading $2\to3$ processes are $\gamma\gamma\to c\bar c(n)gg$ and
$\gamma\gamma\to c\bar c(n)q\bar q$, where $n={}^3\!S_1^{(8)}$,
${}^1\!S_0^{(8)}$, and ${}^3\!P_J^{(8)}$.
The corresponding Feynman diagrams are presented in Figs.~\ref{fig:ccgg}(a)
and \ref{fig:ccqq}, respectively.
From the technical point of view, it is convenient to evaluate the gluon
polarization sum as
$\sum_{\rm pol}\varepsilon_\mu(q)\varepsilon_\nu^*(q)=-g_{\mu\nu}$, at the
expense of allowing for Faddeev-Popov ghosts of the gluon to appear in the
final state; see Fig.~\ref{fig:ccgg}(b).
The corresponding colour-singlet processes with $n={}^3\!S_1^{(1)}$ are
forbidden.
If there is a $gg$ system in the final state, this follows from Furry's
theorem \cite{fur} by observing that the ${}^3\!S_1^{(1)}$ projector
effectively closes the $c$-quark line and acts like a vector coupling and that
the two gluons are then in a colour-singlet state, so that we are dealing with
a closed fermion loop containing five vector couplings.
This was also verified by explicit calculation.
In the case of a $c\bar cq\bar q$ final state, the two quark lines are, at the
order considered, connected by a single gluon, which ensures that the
$c\bar c$ and $q\bar q$ pairs are both in a colour-octet state.

In this work, we calculate for the first time the partonic cross sections of
these $2\to3$ subprocesses.
Requiring the two massless partons accompanying the $c\bar c$ pair to be hard
and isolated, these results can be used to describe experimental data of
$\gamma\gamma\to J/\psi+2j$ if the $J/\psi$ transverse momentum and/or the
dijet invariant mass are sufficiently large, so that the corresponding single-
and double-resolved contributions can be neglected.
Such data were taken at CERN LEP2 with moderate statistics, but a future
$e^+e^-$ collider is bound to deliver a copious amount.
This avenue is taken here.
As a by-product of our analysis, we check previous results for the $2\to2$
processes mentioned above \cite{ma,jap,god}.
In addition, we present the cross section of the partonic subprocess
$\gamma\gamma\to c\bar c\left({}^1\!P_1^{(8)}\right)g$.
Since
$\left\langle{\cal O}_8^{J/\psi}\left({}^1\!P_1\right)\right\rangle$, which
has not yet been extracted from experimental data, is formally of
${\it O}(v^4)$ relative to
$\left\langle{\cal O}_8^{J/\psi}\left({}^3\!S_1\right)\right\rangle$, this
yields a relativistic correction to the cross section of
$\gamma\gamma\to J/\psi+j$.

On the other hand, the $2\to3$ cross sections under consideration here
constitute an essential ingredient for the calculation of the NLO corrections
to the inclusive cross section of $\gamma\gamma\to J/\psi+X$.
In fact, integrating out the phase space of the two massless partons, one 
obtains the real radiative corrections, which suffer from both infrared (IR)
singularities and collinear ones associated with the incoming photon legs.
The latter are factorized, at some factorization scale $M$, and absorbed into
the bare photon PDFs appearing in the appropriate single- and double-resolved
cross sections, so as to render these bare PDFs renormalized.
In this way, the direct, single-resolved, and double-resolved cross sections
all become $M$ dependent.
However, this $M$ dependence cancels in their sum, up to terms beyond NLO.
The IR singularities cancel when the real radiative corrections are combined 
with the virtual ones.
Finally, the ultraviolet (UV) radiative corrections contained in the latter
are removed by renormalizing the couplings, masses, wave-functions, and
non-perturbative MEs appearing in the LO cross section of
$\gamma\gamma\to J/\psi+X$.

The single- and double-resolved cross sections of $\gamma\gamma\to J/\psi+j$
were studied in Refs.~\cite{jap,god}, on the basis of the analytic results of
Refs.~\cite{ber,bkv,ko} and \cite{bai,cho,bkv}, respectively, and found to be
generally more important than the direct one, especially for small values of 
$p_T$.
For the reasons explained above, our analysis is complementary to the ones of
Refs.~\cite{jap,god}.

This paper is organized as follows.
In Section~\ref{sec:two}, we present, in analytic form, the cross sections of
the partonic subprocesses
$\gamma\gamma\to c\bar c\left({}^3\!S_1^{(8)}\right)g$,
$\gamma\gamma\to c\bar c\left({}^3\!S_1^{(1)}\right)\gamma$, and
$\gamma\gamma\to c\bar c\left({}^1\!P_1^{(8)}\right)g$.
Furthermore, we describe the kinematics of the $2\to2$ and $2\to3$ processes.
The analytic results for the latter are too long to be listed.
In Section~\ref{sec:three}, we present our numerical results for the cross 
sections of $\gamma\gamma\to J/\psi+X$, where $X=j,\gamma$, and
$\gamma\gamma\to J/\psi+2j$ appropriate for LEP2, TESLA, and the 
$\gamma\gamma$ collider mode of the latter.
Our conclusions are summarized in Section~\ref{sec:four}.

\section{Details of the calculation}
\label{sec:two}

To start with, we list the cross sections of the partonic subprocesses
$\gamma\gamma\to c\bar c\left({}^3\!S_1^{(8)}\right)g$,
$\gamma\gamma\to c\bar c\left({}^3\!S_1^{(1)}\right)\gamma$, and
$\gamma\gamma\to c\bar c\left({}^1\!P_1^{(8)}\right)g$.
Except for the last case, which we only consider at LO, we work in dimensional
regularization with $D=4-2\epsilon$ space-time dimension and introduce a
't~Hooft mass $\mu$.
In this way, our results can be employed for a future NLO analysis as they
stand.
In the LO evaluation to be performed in Section~\ref{sec:three}, we put
$\epsilon=0$.
We apply the projection method of Ref.~\cite{man}, which is equivalent to the
$D$-dimensional matching procedure of Ref.~\cite{bch}, in order to extract the
short-distance coefficients which multiply the MEs.
However, in order to conform with common standards, we adopt the
normalizations of the MEs from Ref.~\cite{bbl} rather than from
Ref.~\cite{man}.
We find
\begin{eqnarray}
\frac{d\sigma}{dt}
\left(\gamma\gamma\to c\bar c\left({}^3\!S_1^{(8)}\right)g\right)
&=&\frac{1}{4(1-\epsilon)^2}\,\frac{1}{\Gamma(1-\epsilon)}
\left(\frac{4\pi\mu^2s}{tu}\right)^\epsilon
\frac{1}{16\pi s^2}\,(4\pi)^32N_cC_F\alpha_sQ_c^4\alpha^2M
\nonumber\\
&&{}\times
\frac{\left\langle{\cal O}_8^{J/\psi}\left({}^3\!S_1\right)\right\rangle}{n_g}
\,\frac{256}{3-2\epsilon}
\nonumber\\
&&{}\times\frac{(2-5\epsilon)stu(s+t+u)+2(1-\epsilon)^2(s^2t^2+s^2u^2+t^2u^2)}
{(s+t)^2(s+u)^2(t+u)^2},
\label{eq:3s1}
\end{eqnarray}
where $Q_c=2/3$ and $M/2=m_c$ are the fractional electric charge and the mass
of the charm quark, respectively, $s$, $t$, and $u$ are the Mandelstam
variables, the first prefactor stems from the average over the photon
polarization, $N_c=3$, $C_F=(N_c^2-1)/(2N_c)$, and $n_g=N_c^2-1$.
The expression for
$d\sigma/dt
\left(\gamma\gamma\to c\bar c\left({}^3\!S_1^{(1)}\right)\gamma\right)$
emerges from Eq.~(\ref{eq:3s1}) through the substitution \cite{ma}
\begin{equation}
2N_cC_F\alpha_s
\frac{\left\langle{\cal O}_8^{J/\psi}\left({}^3\!S_1\right)\right\rangle}{n_g}
\to Q_c^2\alpha
\left\langle{\cal O}_1^{J/\psi}\left({}^3\!S_1\right)\right\rangle,
\label{eq:sub}
\end{equation}
in contrast to what is stated in Ref.~\cite{jap}.
In the physical limit, $\epsilon=0$, Eq.~(\ref{eq:3s1}) agrees with Eq.~(7)
of Ref.~\cite{ma}, while it disagrees with the corresponding result presented
in Ref.~\cite{jap}.
Our result for
$d\sigma/dt
\left(\gamma\gamma\to c\bar c\left({}^3\!S_1^{(1)}\right)\gamma\right)$
disagrees with Eq.~(7) of Ref.~\cite{jap}.
Furthermore, we obtain
\begin{eqnarray}
\frac{d\sigma}{dt}
\left(\gamma\gamma\to c\bar c\left({}^1\!P_1^{(8)}\right)g\right)
&=&\frac{1}{4}\,\frac{1}{16\pi s^2}\,
\frac{(4\pi)^32N_cC_F\alpha_sQ_c^4\alpha^2}{M}\,
\frac{\left\langle{\cal O}_8^{J/\psi}\left({}^1\!P_1\right)\right\rangle}{n_g}
\nonumber\\
&&{}\times\frac{2048}{3[(s+t)^3(s+u)^3(t+u)^3]}\{-7(stu)^2(s+t+u)
\nonumber\\
&&{}-5stu(s+t+u)^2(st+su+tu)+4stu(s+t+u)^4
\nonumber\\
&&{}-[s^2t^2(s+t)^3+s^2u^2(s+u)^3+t^2u^2(t+u)^3]
\nonumber\\
&&{}+st(s+t)^5+su(s+u)^5+tu(t+u)^5\}.
\label{eq:1p1}
\end{eqnarray}
For definiteness, the non-perturbative MEs in Eqs.~(\ref{eq:3s1}) and
(\ref{eq:1p1}) refer to the $J/\psi$ meson, but these equations are also valid
for other heavy quarkonia.

The respective cross section of $e^+e^-\to e^+e^-J/\psi+X$ is obtained by
convoluting the sum of these partonic cross sections with the photon flux
functions $f_{\gamma/e}(x)$, where $x$ denotes the fraction of the electron or
positron beam energy carried by the bremsstrahlung, beamstrahlung, or laser
photons.
Working in the $e^+e^-$ centre-of-mass (CM) frame and denoting the nominal
$e^+e^-$ energy by $\sqrt S$, the transverse momentum and rapidity of the
$J/\psi$ meson by $p_T$ and $y$, and those of the gluon jet or hard photon by
$p_{T1}$ and $y_1$, we have
\begin{equation}
\frac{d^3\sigma(e^+e^-\to e^+e^-J/\psi+X)}{dp_T^2dydy_1}
=x_+f_{\gamma/e}(x_+)x_-f_{\gamma/e}(x_-)
\sum_n\sum_{a=g,\gamma}
\frac{d\sigma}{dt}(\gamma\gamma\to c\bar c(n)a),
\label{eq:2to2}
\end{equation}
where $x_\pm=[m_T\exp(\pm y)+p_{T1}\exp(\pm y_1)]/\sqrt S$, with
$m_T=\sqrt{p_T^2+M^2}$ and $p_{T1}=p_T$.
The Mandelstam variables are then given by
$s=x_+x_-S$, $t=M^2-x_+\sqrt Sm_T\exp(-y)$, and $u=M^2-x_-\sqrt Sm_T\exp(y)$.
For a given value of $\sqrt S$, the accessible phase space is defined by
\begin{eqnarray}
&&0\le p_T\le \frac{S-M^2}{2\sqrt{S}},\nonumber\\
&&|y|\le\Arcosh\frac{S+M^2}{2\sqrt Sm_T},\nonumber\\
&&-\ln\frac{\sqrt S-m_T\exp(-y)}{p_{T1}}
\le y_1\le\ln\frac{\sqrt S-m_T\exp(y)}{p_{T1}}.
\end{eqnarray}

We now turn to the case where the system $X$ consists of two hadron jets.
Calling the azimuthal angles, transverse momenta, and rapidities of the latter
$\phi_i$, $p_{Ti}$, and $y_i$, with $i=1,2$, we have
\begin{eqnarray}
\frac{d^6\sigma(e^+e^-\to e^+e^-J/\psi+2j)}
{dp_T^2dyd\phi_1dp_{T1}^2dy_1dy_2}
&=&\frac{1}{512\pi^4s^2}
x_+f_{\gamma/e}(x_+)x_-f_{\gamma/e}(x_-)
\nonumber\\
&&{}\times\sum_n\sum_{ab=gg,q\bar q}\overline{|{\cal M}^2|}
(\gamma\gamma\to c\bar c(n)ab),
\end{eqnarray}
where
$x_\pm=[m_T\exp(\pm y)+p_{T1}\exp(\pm y_1)+p_{T2}\exp(\pm y_2)]/
\sqrt S$,
with $p_{T2}={}$\break
$\sqrt{p_T^2+p_{T1}^2+2p_Tp_{T1}\cos\phi_1}$.
The analytic expressions for the various spin-averaged, squared matrix
elements $\overline{|{\cal M}^2|}(\gamma\gamma\to c\bar c(n)ab)$ are too
lengthy to be listed here.
The accessible phase space is now defined by
\begin{eqnarray}
&&0\le p_T\le\frac{S-M^2}{2\sqrt S},
\\
&&|y|\le\Arcosh\frac{S+M^2}{2\sqrt Sm_T},
\nonumber\\
&&0\le\phi_1\le2\pi,
\nonumber\\
&&0\le p_{T1}\le\frac{S+M^2-2\sqrt Sm_T\cosh y}
{2\left(\sqrt{S+m_T^2-2\sqrt Sm_T\cosh y}+p_T\cos\phi_1\right)},
\nonumber\\
&&\left|y_1-\frac{1}{2}
\ln\frac{\sqrt S-m_T\exp(y)}{\sqrt S-m_T\exp(-y)}\right|
\le\Arcosh\frac{S+M^2-2\sqrt Sm_T\cosh y-2p_Tp_{T1}\cos\phi_1}
{2p_{T1}\sqrt{S+m_T^2-2\sqrt Sm_T\cosh y}},
\nonumber\\
&&-\ln\frac{\sqrt S-m_T\exp(-y)-p_{T1}\exp(-y_1)}{p_{T2}}
\le y_2\le\ln\frac{\sqrt S-m_T\exp(y)-p_{T1}\exp(y_1)}{p_{T2}}.
\nonumber
\end{eqnarray}

\section{Numerical analysis}
\label{sec:three}

We are now in a position to explore the phenomenological implications of our
results.
We use $\alpha=1/137.036$ \cite{pdg} and evaluate $\alpha_s^{(n_f)}(\mu)$
from the LO formula taking the number of active quark flavours to be $n_f=3$,
the renormalization scale to be $\mu=m_T$, and the asymptotic scale parameter
to be $\Lambda_{\rm QCD}^{(3)}=145$~MeV, which corresponds to
$\alpha_s^{(5)}(M_Z)=0.1180$ \cite{kkp} if the charm- and bottom-quark
thresholds are chosen to be at $m_c=1.5$~GeV and $m_b=4.5$~GeV \cite{pdg}.
We adopt the nonperturbative $J/\psi$-meson MEs appropriate for the MRST
proton PDFs \cite{mar} from Ref.~\cite{bkl},
$\left\langle{\cal O}_1^{J/\psi}\left({}^3\!S_1\right)\right\rangle
=1.3$~GeV${}^3$,
$\left\langle{\cal O}_8^{J/\psi}\left({}^3\!S_1\right)\right\rangle
=4.4\times10^{-3}$~GeV${}^3$, and
$M_{3.4}^{J/\psi}=8.7\times10^{-2}$~GeV${}^3$,
together with the multiplicity relation
\begin{equation}
\left\langle{\cal O}_8^{J/\psi}\left({}^3\!P_J\right)\right\rangle
=(2J+1)\left\langle{\cal O}_8^{J/\psi}\left({}^3\!P_0\right)\right\rangle,
\end{equation}
which follows from heavy-quark spin symmetry.
In want of more specific information, we democratically split the linear
combination
\begin{equation}
M_r=\left\langle{\cal O}_8^{J/\psi}\left({}^1\!S_0\right)\right\rangle
+\frac{r}{m_c^2}
\left\langle{\cal O}_8^{J/\psi}\left({}^3\!P_0\right)\right\rangle
\end{equation}
as
$\left\langle{\cal O}_8^{J/\psi}\left({}^1\!S_0\right)\right\rangle
=\left(r/m_c^2\right)
\left\langle{\cal O}_8^{J/\psi}\left({}^3\!P_0\right)\right\rangle
=M_r/2$.

In $\gamma\gamma$ collisions at LEP2, the scattered electrons and positrons
are usually antitagged, a typical value for the maximum scattering angle being
$\theta_{\rm max}=33$~mrad \cite{abb}.
The energy spectrum of the bremsstrahlung photons is then well described in
the Weizs\"acker-Williams approximation (WWA) \cite{wei} by Eq.~(27) of
Ref.~\cite{fri}.
Values as small as $\theta_{\rm max}=20$~mrad should be feasible at TESLA
\cite{des}.
As already mentioned in the Introduction, at high-energy $e^+e^-$ LCs, hard
photons also arise from beamstrahlung.
The energy spectrum of these beamstrahlung photons is approximately described
by Eq.~(2.14) of Ref.~\cite{che}.
It is controlled the beamstrahlung parameter $\Upsilon$, which is estimated to
be $\Upsilon=0.040$ for TESLA.
We coherently superimpose the WWA and beamstrahlung spectra.
Finally, in the case of a $\gamma\gamma$ collider, the energy spectrum of the
back-scattered laser photons is given by Eq.~(6a) of Ref.~\cite{gin}.
It depends on the parameter $\kappa=s_{e\gamma}/m_e^2-1$, where
$\sqrt{s_{e\gamma}}$ is the CM energy of the charged lepton and the laser 
photon, and it extends up to $x_{\rm max}=\kappa/(\kappa+1)$, where $x$ is the
energy of the back-scattered photons in units of $\sqrt S/2$.
The optimal value of $\kappa$ is $\kappa=2\left(1+\sqrt2\right)\approx4.83$
\cite{tel}; for larger values of $\kappa$, $e^+e^-$ pairs would be created in
the collisions of laser and back-scattered photons.
Representative $e^+e^-$ CM energies of LEP2 and TESLA are $\sqrt S=189$~GeV 
and 500~GeV.

\begin{figure}[ht]
\begin{center}
\epsfig{figure=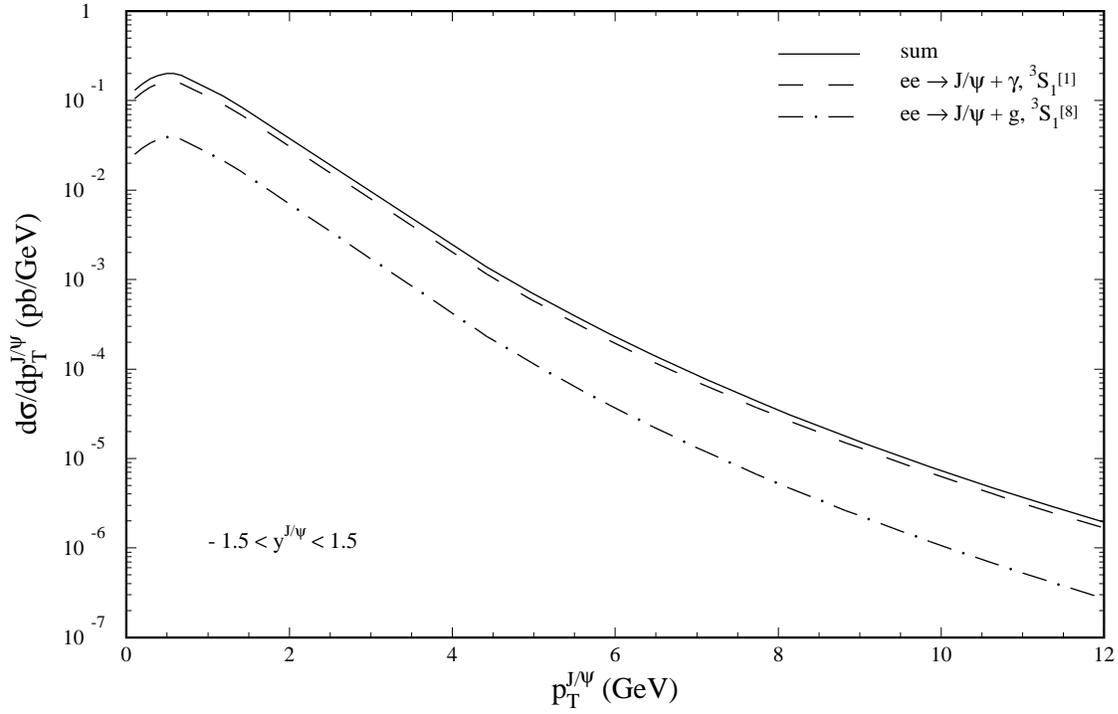,width=\textwidth}
\caption{Transverse-momentum distribution $d\sigma/dp_T$, integrated over
rapidity interval $|y|<1.5$, of $\gamma\gamma\to J/\psi+X$, where $X$
represents a gluon jet or a prompt photon, via bremsstrahlung at LEP2.
The contributions corresponding to these two final states are also shown
separately.}
\label{fig:pccxl}
\end{center}
\end{figure}

\begin{figure}[ht]
\begin{center}
\epsfig{figure=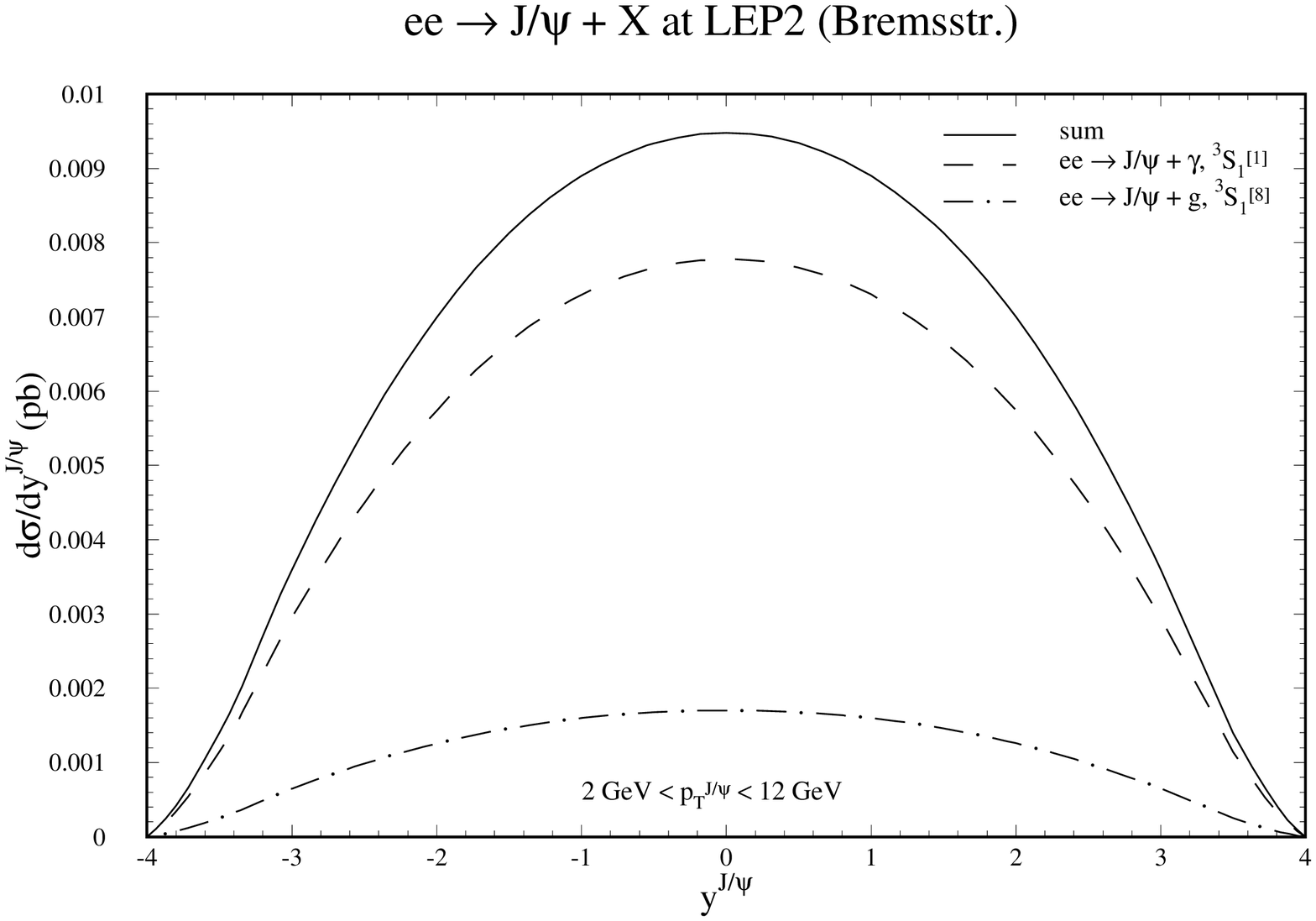,width=\textwidth}
\caption{Rapidity distribution $d\sigma/dy$, integrated over 
transverse-momentum interval $2<p_T<12$~GeV, of $\gamma\gamma\to J/\psi+X$,
where $X$ represents a gluon jet or a prompt photon, via bremsstrahlung at
LEP2.
The contributions corresponding to these two final states are also shown
separately.}
\label{fig:yccxl}
\end{center}
\end{figure}

\begin{figure}[ht]
\begin{center}
\epsfig{figure=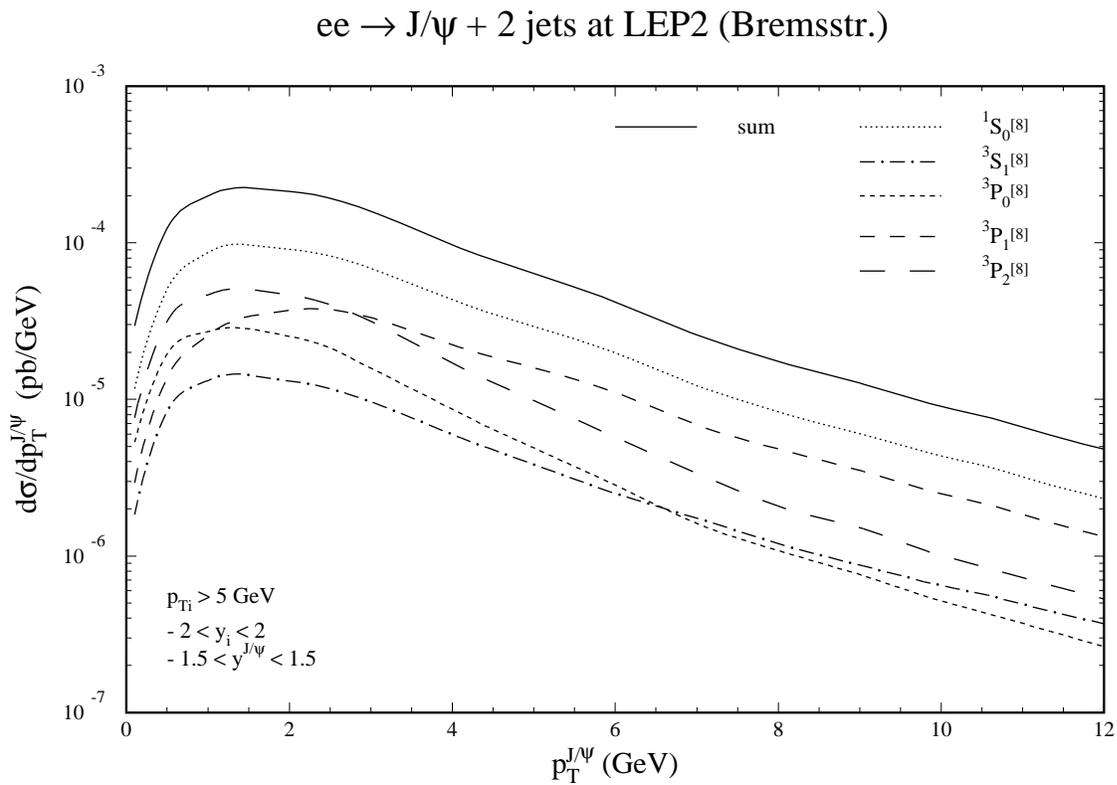,width=\textwidth}
\caption{Transverse-momentum distribution $d\sigma/dp_T$ of
$\gamma\gamma\to J/\psi+jj$ via bremsstrahlung at LEP2.
The contributions due to the various colour-octet channels are also shown
separately.}
\label{fig:pccjjl}
\end{center}
\end{figure}

\begin{figure}[ht]
\begin{center}
\epsfig{figure=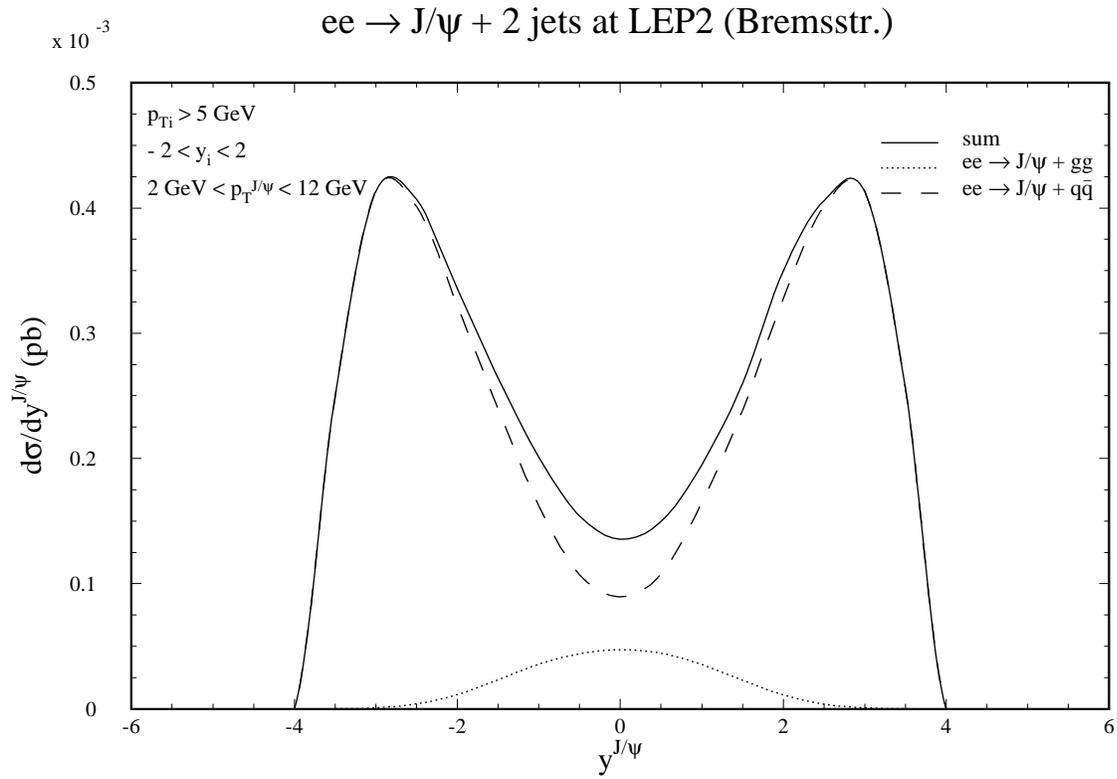,width=\textwidth}
\caption{Rapidity distribution $d\sigma/dy$ of $\gamma\gamma\to J/\psi+jj$ via 
bremsstrahlung at LEP2.
The contributions due to quark and gluon dijets are also shown separately.}
\label{fig:yccjjl}
\end{center}
\end{figure}

\begin{figure}[ht]
\begin{center}
\epsfig{figure=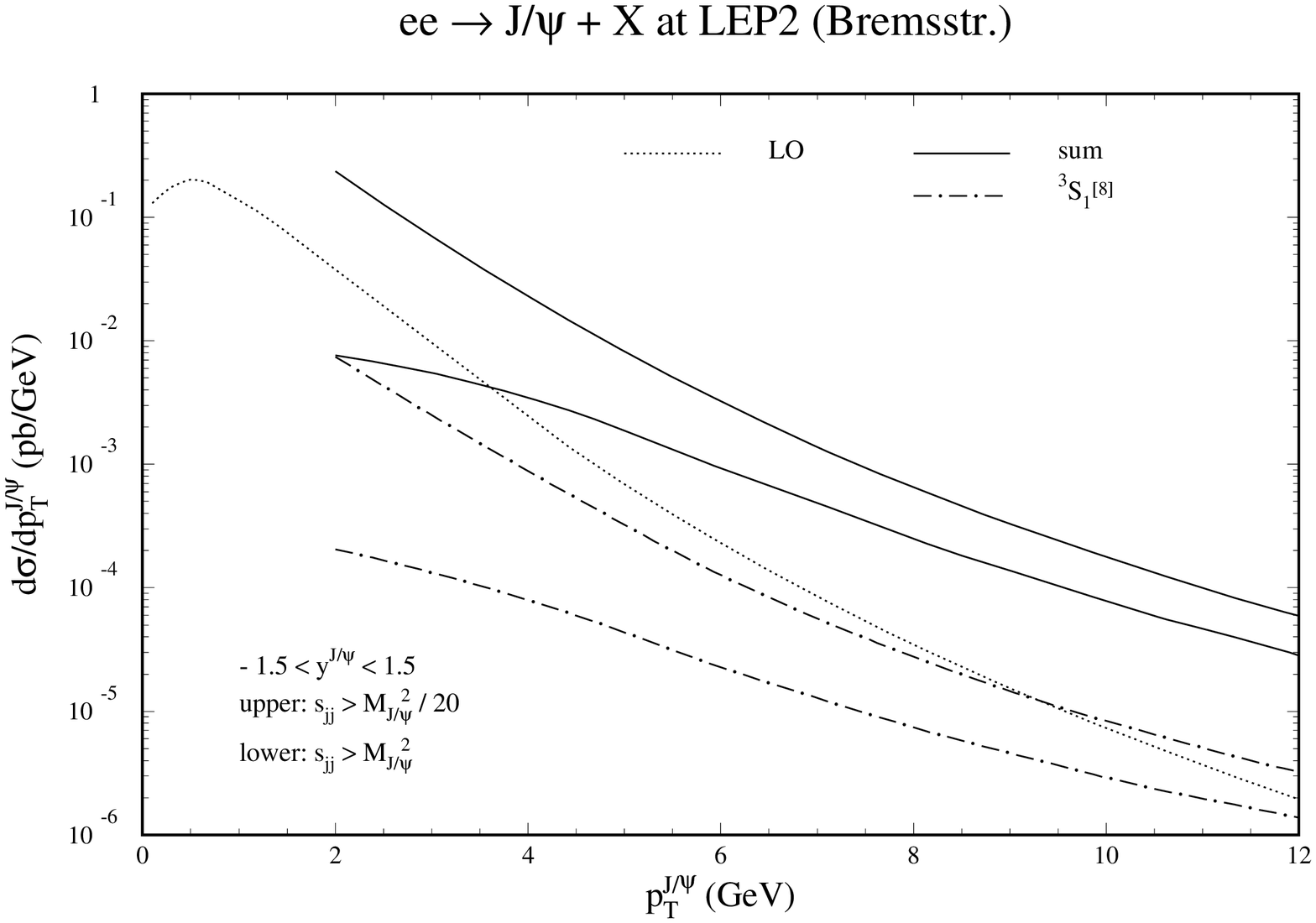,width=\textwidth}
\caption{Transverse-momentum distribution $d\sigma/dp_T$ of
$\gamma\gamma\to J/\psi+X$ via bremsstrahlung at LEP2.
The sum of the LO contributions for $X=j$ and $X=\gamma$ is compared with the
$2\to3$ part of the NLO contribution for dijet invariant mass $s_{jj}>M^2$ and
$M^2/20$.
For comparison, also the ${}^3\!S_1^{(8)}$-channel contributions to the latter
are shown.}
\label{fig:pccjjcut}
\end{center}
\end{figure}

In Figs.~\ref{fig:pccxl}--\ref{fig:pccjjcut}, we quantitatively investigate
the cross section of inclusive $J/\psi$-meson production in the collisions
of two bremsstrahlung photons at LEP2.
In Figs.~\ref{fig:pccxl} and \ref{fig:yccxl}, we present for
$\gamma\gamma\to J/\psi+X$, where $X$ represents a gluon jet (dot-dashed
lines) or a prompt photon (dashed lines), the distributions in transverse
momentum $p_T$ and rapidity $y$ of the $J/\psi$ meson, respectively.
The solid lines refer to the sum of the contributions for $X=j$ and 
$X=\gamma$.
The cuts on $p_T$ and $y$ are adopted from Ref.~\cite{abb}.
In the case $X=j$, the leading contribution comes from the ${}^3\!S_1^{(8)}$
channel.
The contribution from the ${}^1\!P_1^{(8)}$ channel is suppressed for the
reason explained in the Introduction.
Furthermore, the value of
$\left\langle{\cal O}_8^{J/\psi}\left({}^1\!P_1\right)\right\rangle$ is not
yet available.
Therefore, this contribution is not included in Figs.~\ref{fig:pccxl} and 
\ref{fig:yccxl}.
The distributions for $X=j$ and $X=\gamma$ only differ by the overall 
normalization, as is evident from Eq.~(\ref{eq:sub}).
Obviously, the coupling suppression of the result for $X=\gamma$ is less
substantial than the suppression of the result for $X=j$ by the fact that
$\left\langle{\cal O}_8^{J/\psi}\left({}^3\!S_1\right)\right\rangle$ is of
order $v^4$ relative to
$\left\langle{\cal O}_1^{J/\psi}\left({}^3\!S_1\right)\right\rangle$.
Since we assumed the same experimental set-up for the electron and positron
beams, the $y$ distributions displayed in Fig.~\ref{fig:yccxl} are all
symmetric about $y=0$.
Figure~\ref{fig:pccxl} is similar to Fig.~1 of Ref.~\cite{god}, which we are 
able to reproduce, adopting the input parameters specified in that reference.
We note in passing that Ref.~\cite{god}, too, disagrees with Ref.~\cite{jap}
for $X=j$ and $X=\gamma$; see the statement contained in Ref.~[21] of
Ref.~\cite{god}.

In Figs.~\ref{fig:pccjjl} and \ref{fig:yccjjl}, we study the $p_T$ and $y$
distributions of $\gamma\gamma\to J/\psi+jj$, respectively, again imposing the
cuts of Ref.~\cite{abb}.
In addition, following Ref.~\cite{wen}, we require for the two jets to have
transverse momenta $p_{Ti}>5$~GeV and rapidities $|y_i|<2$ ($i=1,2$) and to be
separated by $\Delta R=\sqrt{(y_1-y_2)^2+(\phi_1-\phi_2)^2}>1$ according to
the $k_T$-clustering algorithm \cite{cat}.
While no actual clustering is performed, the separation of the two jets is
necessary to avoid collinear singularities in the final state.
In Fig.~\ref{fig:pccjjl}, we analyze the relative importance of the various
colour-octet channels.
We observe that the ${}^1\!S_0^{(8)}$ channel is most important.
As is familiar from inclusive $J/\psi$-meson hadroproduction at the Tevatron
\cite{cho}, the ${}^1\!S_0^{(8)}$ and ${}^3\!P_J^{(8)}$ channels have very
similar $p_T$ dependences for $p_T\agt5$~GeV.
By the same token, this implies that the theoretical uncertainty related to 
the lack of information on how $M_r$ breaks into
$\left\langle{\cal O}_8^{J/\psi}\left({}^1\!S_0\right)\right\rangle$ and
$\left\langle{\cal O}_8^{J/\psi}\left({}^3\!P_0\right)\right\rangle$ is
modest.
On the other hand, the ${}^3\!S_1^{(8)}$-channel contribution falls off less
rapidly as $p_T$ increases.
The suppression of the ${}^3\!S_1^{(8)}$ contribution relative to the
combination of the ${}^1\!S_0^{(8)}$ and ${}^3\!P_J^{(8)}$ ones may be traced
to the fact that $M_r$ is almost a factor of 20 larger than
$\left\langle{\cal O}_8^{J/\psi}\left({}^3\!S_1\right)\right\rangle$ 
\cite{bkl}.
The total $y$ distribution shown in Fig.~\ref{fig:yccjjl} (solid line)
exhibits a marked minimum at $y=0$.
Its decomposition into the contributions where the dijets are of $gg$ (dotted
line) and $q\bar q$ (dashed line) origin clarifies that this minimum stems
from the latter contribution.
This may be understood by observing that the $q$ and $\bar q$ quarks may be 
created from the splitting of the incident photons, in which case they are
dominantly collinear to the mother photons.
On the other hand, the final-state gluons are either both directly radiated
off the heavy-quark line or emerge through the $1\to2$ splitting of a virtual
gluon that is radiated off the heavy-quark line, so that the resulting $y$ 
distribution is expected to have a shape similar to the one of
$\gamma\gamma\to J/\psi+j$ studied in Fig.~\ref{fig:yccxl}, which has a 
maximum at $y=0$.

In Fig.~\ref{fig:pccjjcut}, we make an attempt to obtain a first hint at the
size of the NLO correction to the cross section of $\gamma\gamma\to J/\psi+X$,
where $X$ is a gluon jet or a prompt photon.
To this end, we compare the LO $p_T$ distribution of
$\gamma\gamma\to J/\psi+X$ (dotted line), taken from Fig.~\ref{fig:pccxl},
with the one of $\gamma\gamma\to J/\psi+jj$ after integrating the dijet 
invariant mass $\sqrt{s_{jj}}$ over all kinematically allowed values with
$s_{jj}>M^2$ (lower solid line) and $M^2/20$ (upper solid line).
A similar strategy was adopted in Ref.~\cite{pet} to estimate the NLO
correction to $J/\psi$-meson hadroproduction via the ${}^3\!S_1^{(1)}$ channel
at finite values of $p_T$.
The dependence on the $s_{jj}$ lower cutoff would be compensated in the 
full NLO result by the virtual and soft real corrections, which are not yet 
available.
However, from Fig.~\ref{fig:pccjjcut} we learn that this dependence is reduced
as the value of $p_T$ increases.
Moreover, the missing part of the NLO contribution should have a similar 
$p_T$ dependence as the LO result, and it should, therefore, be suppressed
relative to the available part, presented in Fig.~\ref{fig:pccjjcut}, at large
values of $p_T$.
Consequently, we expect the NLO correction factor to be significantly larger 
than unity in the high-$p_T$ regime.
Of course, a solid statement can only be made on the basis of a complete NLO
analysis.
For comparison, we also indicate in Fig.~\ref{fig:pccjjcut} the
${}^3\!S_1^{(8)}$-channel portions of the contributions represented by the 
solid lines.
They are scaled down by a factor of 20 to 30, as is naively expected from the
$\left\langle{\cal O}_8^{J/\psi}\left({}^3\!S_1\right)\right\rangle$ to
$M_r$ ratio \cite{bkl}.
Strictly speaking, $\gamma\gamma\to J/\psi+jj$ contributes to the real QCD
correction to $\gamma\gamma\to J/\psi+j$, but not to
$\gamma\gamma\to J/\psi+\gamma$.
However, the NLO correction to the latter process is suppressed by the factor
$\alpha/\alpha_s$ relative to the NLO correction under consideration here, and
it can thus be safely neglected.

\begin{figure}[ht]
\begin{center}
\epsfig{figure=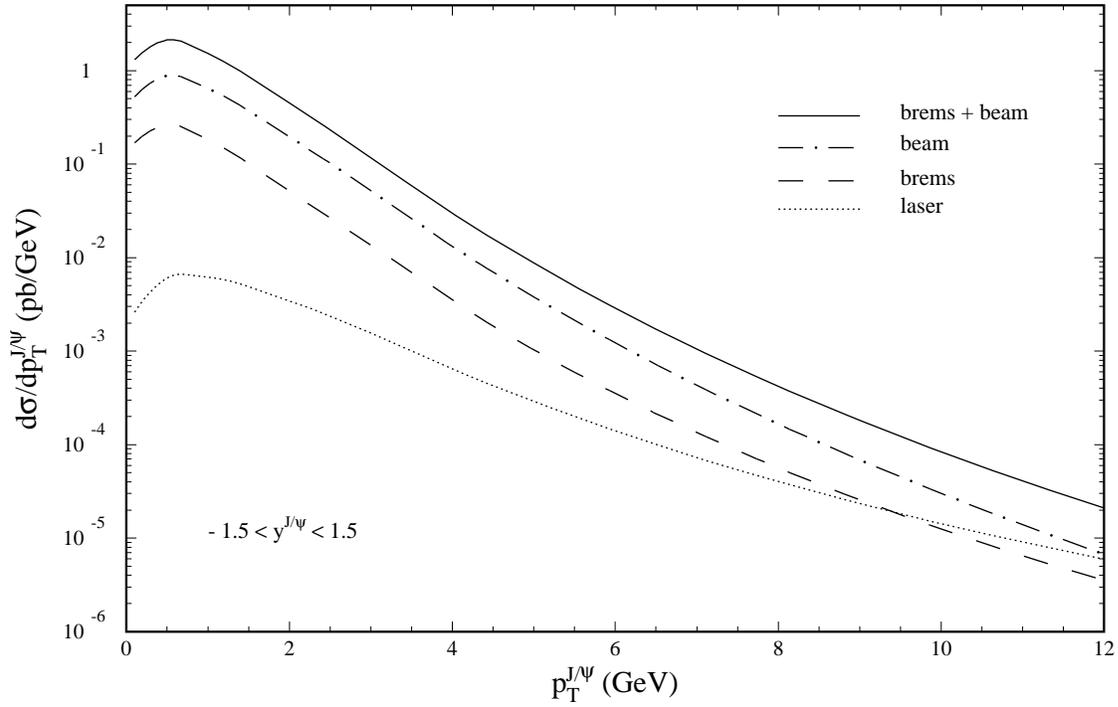,width=\textwidth}
\caption{Transverse-momentum distribution $d\sigma/dp_T$, integrated over
rapidity interval $|y|<1.5$, of $\gamma\gamma\to J/\psi+X$, where $X$
represents a gluon jet or a prompt photon, via bremsstrahlung, beamstrahlung,
their coherent superposition, and laser back-scattering at TESLA.}
\label{fig:pccxt}
\end{center}
\end{figure}

\begin{figure}[ht]
\begin{center}
\epsfig{figure=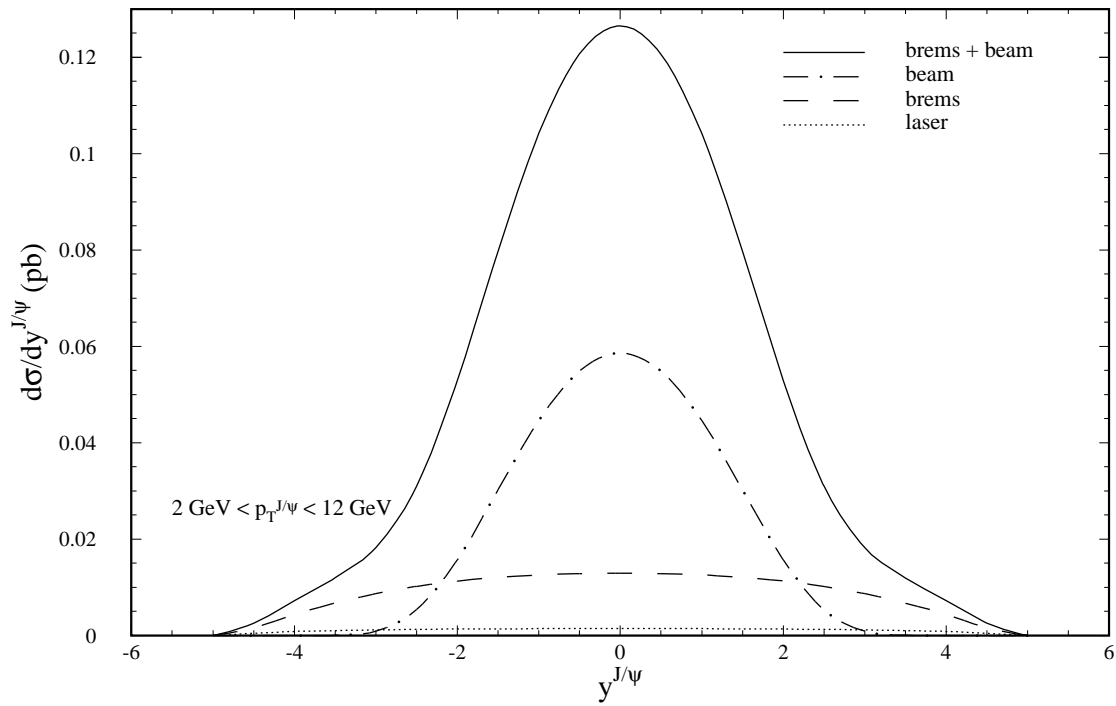,width=\textwidth}
\caption{Rapidity distribution $d\sigma/dy$, integrated over
transverse-momentum interval $2<p_T<12$~GeV, of $\gamma\gamma\to J/\psi+X$,
where $X$ represents a gluon jet or a prompt photon, via bremsstrahlung,
beamstrahlung, their coherent superposition, and laser back-scattering at
TESLA.}
\label{fig:yccxt}
\end{center}
\end{figure}

\begin{figure}[ht]
\begin{center}
\epsfig{figure=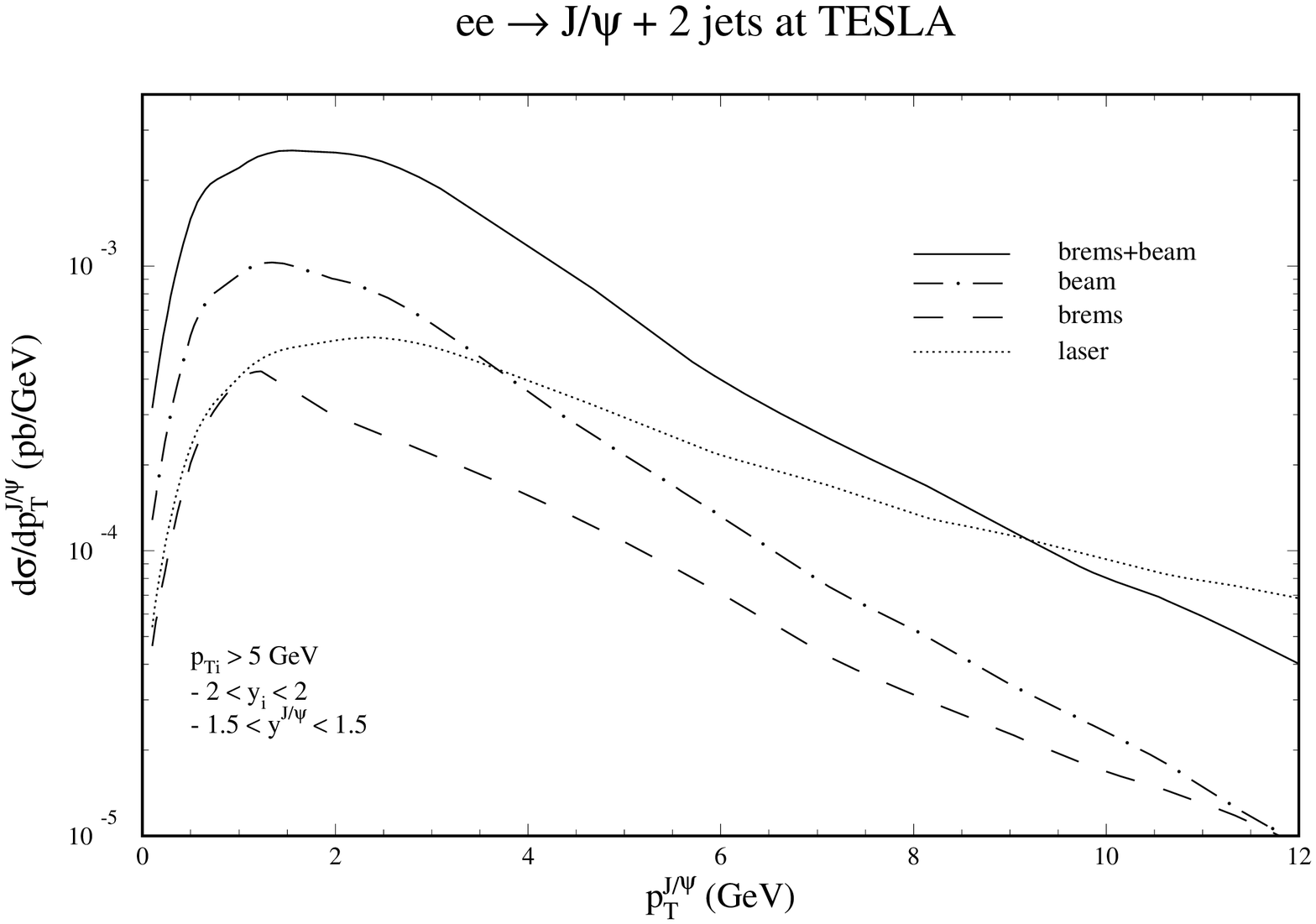,width=\textwidth}
\caption{Transverse-momentum distribution $d\sigma/dp_T$ of
$\gamma\gamma\to J/\psi+jj$ via bremsstrahlung, beamstrahlung, their coherent
superposition, and laser back-scattering at TESLA.}
\label{fig:pccjjt}
\end{center}
\end{figure}

\begin{figure}[ht]
\begin{center}
\epsfig{figure=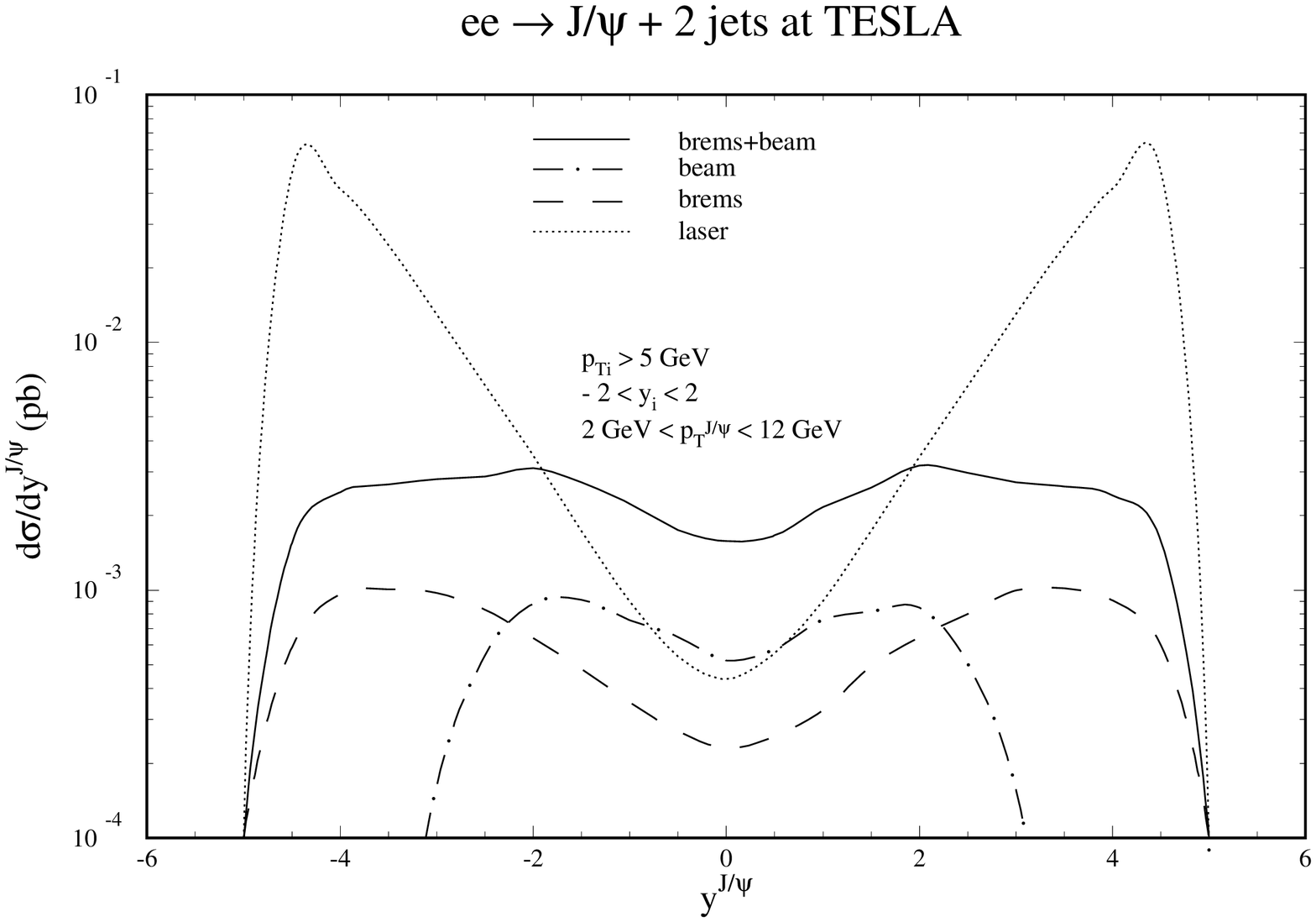,width=\textwidth}
\caption{Rapidity distribution $d\sigma/dy$ of $\gamma\gamma\to J/\psi+jj$ via
bremsstrahlung, beamstrahlung, their coherent superposition, and laser
back-scattering at TESLA.}
\label{fig:yccjjt}
\end{center}
\end{figure}

We now turn to $\gamma\gamma$ collisions at TESLA.
As explained above, in the $e^+e^-$ mode, the photons originate from
bremsstrahlung and beamstrahlung, while in the $\gamma\gamma$ mode, they
arise from the back-scattering of laser light on the incident electron and
positron beams.
In Figs.~\ref{fig:pccxt}--\ref{fig:yccjjt}, the contributions due to
bremsstrahlung (dashed lines), beamstrahlung (dot-dashed lines), their
coherent superposition (solid lines), and Compton scattering (dotted lines)
are shown separately.
We apply the same cuts on $p_T$, $y$, $p_{Ti}$, and $y_i$ ($i=1,2$) as in the
LEP2 case.
The $p_T$ and $y$ distributions of $\gamma\gamma\to J/\psi+X$, where it is
summed over $X=j$ and $X=\gamma$, are presented in Figs.~\ref{fig:pccxt} and
\ref{fig:yccxt}, respectively.
From Eq.~(\ref{eq:2to2}) we know that small values of $p_T$ typically
correspond to small values of $x$ and vice versa.
Since the bremsstrahlung and beamstrahlung spectra are peaked at $x=0$, while
the Compton spectrum is peaked at $x=x_{\rm max}$, it hence follows that the
bremsstrahlung and beamstrahlung contributions significantly overshoot the
Compton one in the small-$p_T$ regime, while the latter wins out at large
values of $p_T$, a feature which so far has gone unnoticed in the literature.
Owing to the rapid fall-off of the $p_T$ distributions in
Fig.~\ref{fig:pccxt}, the corresponding $y$ distributions in
Fig.~\ref{fig:yccxt} receive their bulk contributions from the small-$p_T$
regime.
This explains why the Compton contribution is greatly suppressed relative to
the bremsstrahlung and beamstrahlung ones.
The beamstrahlung contribution exhibits a prominent peak about $y=0$ and drops
off at $|y|\approx3$, while the bremsstrahlung one has a flatter shape and
dominates for $|y|\agt2.2$.

The $p_T$ and $y$ distributions of $\gamma\gamma\to J/\psi+jj$, with all the
leading colour-octet channels included, are shown in Figs.~\ref{fig:pccjjt}
and \ref{fig:yccjjt}, respectively.
Comparing Fig.~\ref{fig:pccjjt} with Fig.~\ref{fig:pccxt}, we observe that now
the Compton contribution starts to exceed the one due to bremsstrahlung and
beamstrahlung already at $p_T\approx9$~GeV.
This may be understood by observing that it is now possible for $x$ to take
large values at small values of $p_T$ if the dijet invariant mass is
sufficiently large.
As in Fig.~\ref{fig:yccjjl}, the $y$ distributions are symmetric about $y=0$
and exhibit a local minimum there.
This feature is particularly pronounced for the Compton contribution.
As in Fig.~\ref{fig:yccxt}, the beamstrahlung contribution exceeds the
bremsstrahlung one in the central $y$ region and only extends out to
$|y|\approx3$, while the bremsstrahlung one dominates for $|y|\agt2.2$.

\section{Conclusions}
\label{sec:four}

We calculated the cross section of $\gamma\gamma\to J/\psi+jj$ in direct
photoproduction at LO in the NRQCD factorization formalism and provided
theoretical predictions for the $J/\psi$-meson $p_T$ and $y$ distributions in
$\gamma\gamma$ collisions via initial-state bremsstrahlung at LEP2 and via
bremsstrahlung and beamstrahlung or laser back-scattering at TESLA.
We also performed a similar study for $\gamma\gamma\to J/\psi+X$, where $X$
represents a gluon jet or a prompt photon, and compared our results with the
literature \cite{ma,jap,god}.
We found agreement with Refs.~\cite{ma,god}, but disagreement with
Ref.~\cite{jap} for both $X=j$ and $X=\gamma$.
The contributions due to single-resolved and double-resolved photoproduction
are expected to be suppressed if the $J/\psi$-meson transverse momentum and/or
the dijet invariant mass are large as compared to the $J/\psi$-meson mass.
In the CSM, only $\gamma\gamma\to J/\psi+\gamma$ can happen in direct 
photoproduction.
Experimental observation of $\gamma\gamma\to J/\psi+j$ or
$\gamma\gamma\to J/\psi+jj$ with the predicted cross sections would provide
evidence for the existence of colour-octet processes in nature.
An interesting feature of $\gamma\gamma\to J/\psi+jj$ is the appearance of a
marked minimum of the $y$ distribution at $y=0$.
The cross section of $\gamma\gamma\to J/\psi+jj$ provides an essential
ingredient for the calculation of the NLO correction to the one of
$\gamma\gamma\to J/\psi+X$.
The virtual and soft real corrections remain to be calculated in order to 
obtain the full NLO correction factor.
Our analysis indicates that the latter is likely to be significantly larger
than unity at large values of $p_T$.

\begin{table}[ht]
\begin{center}
\caption{Integrated cross sections (in fb) of $\gamma\gamma\to J/\psi+X$, with
$X=j,\gamma,jj$, via direct photoproduction in $\gamma\gamma$ collisions at
LEP2, TESLA, and its $\gamma\gamma$ option.
The applied cuts are $2<p_T<12$~GeV and $|y|<1.5$ for the $J/\psi$ meson and
$p_{Ti}>5$~GeV, $|y_i|<2$ ($i=1,2$), and $\Delta R>1$ for the dijet system.}
\label{tab:tot}
\medskip
\begin{tabular}{|c|ccc|} \hline\hline
Experiment & $\sigma(J/\psi+j)$ & $\sigma(J/\psi+\gamma)$ &
$\sigma(J/\psi+jj)$ \\
\hline
LEP2 & 4.9 & 59 & 0.74 \\
TESLA $e^+e^-$ & 22 & 270 & 3.5 \\
TESLA $\gamma\gamma$ & 0.55 & 6.2 & 2.4 \\
\hline\hline
\end{tabular}
\end{center}
\end{table}

At $e^+e^-$ colliders, the $J/\psi$ meson can be easily detected through its
decay to a $\mu^+\mu^-$ pair, with branching fraction $(5.88\pm0.10)\%$
\cite{pdg}.
Unfortunately, the LEP2 experiments have not yet extracted from their recorded
data the cross section of inclusive $J/\psi$-meson production in
$\gamma\gamma$ collisions.
The OPAL analysis of $\gamma\gamma\to2j+X$ is based on an integrated
luminosity of 384~pb${}^{-1}$ \cite{wen}.
Assuming that the other LEP2 experiments, ALEPH, DELPHI, and L3, have data
samples of similar sizes and folding in the $J/\psi\to\mu^+\mu^-$ branching
fraction, we conclude that a $J/\psi$-meson production cross section of 1~pb
translates into approximately 90 signal events at LEP2.
The design luminosities for the $e^+e^-$ and $\gamma\gamma$ modes of TESLA,
with $\sqrt S=500$~GeV, are $3.4\times10^{34}$~cm${}^{-2}$s${}^{-1}$ and
$0.6\times10^{34}$~cm${}^{-2}$s${}^{-1}$ \cite{tdr}, respectively, which
corresponds to 340~fb${}^{-1}$ and 60~fb${}^{-1}$ per year.
Thus, a $J/\psi$-meson production cross section of 1~fb yields about 20 and
3.5 signal events at TESLA operating in the $e^+e^-$ and $\gamma\gamma$ modes,
respectively.
In Table~\ref{tab:tot}, we list the integrated cross sections of
$\gamma\gamma\to J/\psi+X$, with $X=j,\gamma,jj$, via direct photoproduction
in $\gamma\gamma$ collisions at LEP2, TESLA, and its $\gamma\gamma$ option.
The applied cuts are $2<p_T<12$~GeV and $|y|<1.5$ for the $J/\psi$ meson and
$p_{Ti}>5$~GeV, $|y_i|<2$ ($i=1,2$), and $\Delta R>1$ for the dijet system.
The respective numbers of expected signal events emerge through multiplication
of the entries in Table~\ref{tab:tot} with the cross section to event number
conversion factors quoted above.

\bigskip
\noindent
{\bf Acknowledgements}
\smallskip

\noindent
We thank J. Lee for a helpful comment on Eq.~(\ref{eq:sub}).
We are grateful to S. S\"oldner-Rembold for a useful communication concerning
the prospects of measuring the cross section of $J/\psi$ inclusive production
in $\gamma\gamma$ collisions at LEP2.
This work was supported in part by the Deutsche Forschungsgemeinschaft through
Grants No.\ KL~1266/1-1 and No.\ KN~365/1-1, by the Bundesministerium f\"ur
Bildung und Forschung through Grant No.\ 05~HT9GUA~3, by the European
Commission through the Research Training Network {\it Quantum Chromodynamics
and the Deep Structure of Elementary Particles} under Contract No.\
ERBFMRX-CT98-0194, and by Sun Microsystems through Academic Equipment Grant
No.~EDUD-7832-000332-GER.

\newpage

\newpage

\newpage

\newpage

\newpage

\newpage

\newpage

\newpage

\newpage

\newpage

\newpage

\newpage

\end{document}